\documentclass{egpubl}
\usepackage{egsr2020}

\usepackage[T1]{fontenc}
\usepackage{dfadobe}  

\usepackage{cite}  % comment out for biblatex with backend=biber
\BibtexOrBiblatex
\electronicVersion
\PrintedOrElectronic
\ifpdf \usepackage[pdftex]{graphicx} \pdfcompresslevel=9
\else \usepackage[dvips]{graphicx} \fi

\usepackage{egweblnk} 
\usepackage{amsmath,bm}
\usepackage{multirow}
\usepackage{makecell}
\usepackage{xcolor}
\usepackage{colortbl}

\title[Deep Photon Mapping]%
      {Deep Photon Mapping}

\author[Zhu et al.]
{
Shilin Zhu$^*$ \quad Zexiang Xu$^*$ \quad Henrik Wann Jensen \quad Hao Su \quad Ravi Ramamoorthi\\
University of California San Diego, United States\\
\{shz338, zexiangxu, haosu\}@eng.ucsd.edu, \{henrik, ravir\}@cs.ucsd.edu
}

\newcommand{\Comment}[1]{}
\newcommand{\Net}{\Phi}
\newcommand{\PhotonNN}{K}
\newcommand{\Scenea}{\textsc{Glass egg}}
\newcommand{\Sceneaa}{\textsc{Glass egg2}}
\newcommand{\Sceneb}{\textsc{Red wine}}
\newcommand{\Scenec}{\textsc{Rings}}
\newcommand{\Scened}{\textsc{Water pool1}}
\newcommand{\Scenee}{\textsc{Water pool2}}
\newcommand{\Scenef}{\textsc{Torus}}
\newcommand{\Sceneg}{\textsc{Dragon}}

\newcommand{\red}[1]{\textcolor{red}{#1}}

\newcommand{\blue}[1]{\textcolor{blue}{#1}}

\newcommand{\boldstart}[1]{\noindent\textbf{#1}}
\newcommand{\boldstartspace}[1]{\vspace{0.1in}\noindent\textbf{#1}}

\begin{document}

%\teaser{
% \includegraphics[width=\linewidth]{eg_new}
% \centering
%  \caption{New EG Logo}
%\label{fig:teaser}
%}

\teaser{
    \centering
    \vspace{-6mm}
    \includegraphics[width=\linewidth]{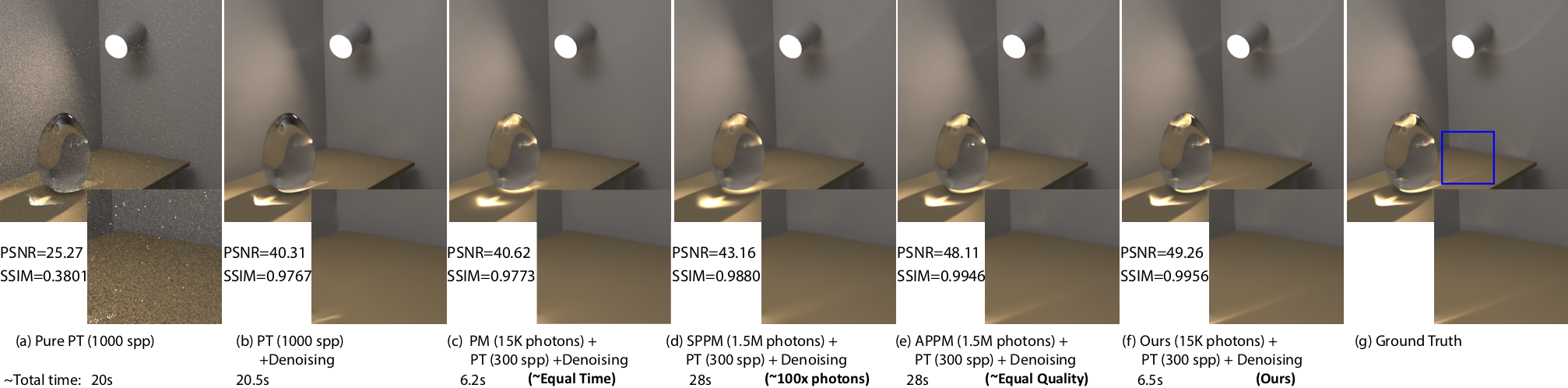}
    \caption{
      We present a novel learning-based photon mapping (PM) method that can be used to synthesize photorealistic images (f) 
      with detailed caustics (shown and compared in the insets) from very sparse photons for scenes with complex diffuse-specular interactions.
      In particular, we use our method with only 15k photons ($\sim$0.06 photons per pixel) to compute accurate global illumination for light-specular paths. 
      We use path tracing (PT) with a moderate number (300) of samples per pixel (spp) to compute the other paths and apply the Optix learning-based denoiser (based on \cite{chaitanya2017interactive}) to remove the Monte Carlo (MC) noise.
      In contrast, pure PT leads to noisy results lacking focused caustics (a) even with 1000 spp that is significantly more than our photon and path samples.
      While this noise can be mitigated using a learning-based denoiser, this introduces artifacts and cannot recover the caustics (b).
      Combining PT and standard PM \cite{jensen1996global} with 15k photons, and then denoising (c), avoids these artifacts but still does not reconstruct caustics accurately from such low photon counts.
      While providing 1.5M photons (this is 100 times the number of photons our method uses) and applying the advanced stochastic progressive PM (SPPM) \cite{hachisuka2010progressive} enables a more accurate result (d), it is still slightly worse than ours.
      In contrast, our result (f) accurately reproduces the caustic effects in the global illumination, as compared to the ground truth (g), with significantly fewer samples.
      Ours is comparable with (if not better than) the result from adaptive progressive PM (APPM) \cite{kaplanyan2013adaptive} with 100 times the number of photons (e).
     }
    \label{fig:teaser}
}

\maketitle
%-------------------------------------------------------------------------
\begin{abstract}
    Recently, deep learning-based denoising approaches have led to dramatic improvements in low sample-count Monte Carlo rendering.  
    These approaches are aimed at path tracing, which is not ideal for simulating challenging light transport effects like caustics, where photon mapping is the method of choice.  
    However, photon mapping requires very large numbers of traced photons to achieve high-quality reconstructions.  
    In this paper, we develop the first deep learning-based method for particle-based rendering, and specifically focus on photon density estimation, the core of all particle-based methods.  
    We train a novel deep neural network to predict a kernel function to aggregate photon contributions at shading points. 
    Our network encodes individual photons into per-photon features, aggregates them in the neighborhood of a shading point to construct a photon local context vector, and infers a kernel function from the per-photon and photon local context features.
    This network is easy to incorporate in many previous photon mapping methods (by simply swapping the kernel density estimator) and can produce high-quality reconstructions of complex global illumination effects like caustics with an order of magnitude fewer photons compared to previous photon mapping methods.
%The tool at \url{http://dl.acm.org/ccs.cfm} can be used to generate
% CCS codes.
%Example:
\Comment{

}

\end{abstract}  
%-------------------------------------------------------------------------

\section{Introduction}
\label{sec:intro}
Computing global illumination is crucial for photorealistic image synthesis.
Ray tracing-based methods have been widely used to simulate complex light transport effects with global illumination in film, animation, video game and other industrial fields. 
The most successful approaches are based on either Monte Carlo (MC) integration, like path tracing \cite{kajiya1986rendering,veach1997robust}, or particle density estimation, like photon mapping \cite{jensen1996global}.
Photon mapping techniques are able to efficiently simulate caustics and other challenging light transport effects,
% including those that involve SDS (specular-diffuse-specular) paths, 
which are very hard and even impossible for pure Monte Carlo-based methods to simulate.

In general, both MC-based and particle-based methods require numerous samples to render noise-free images, and are thus computationally expensive.
Recently, significant progress has been made in denoising MC images rendered with low sample counts using deep learning techniques \cite{chaitanya2017interactive,bako2017kernel}.
However, there is relatively little work in particle-based methods for low-sample reconstruction and current photon mapping techniques still require a very large number of traced photons to achieve accurate, artifact-free radiance estimation.

%Our goal is to make particle-based rendering more efficient to enable highly efficient global illumination computation.
{\em We present the first deep learning-based approach for particle-based rendering} that enables efficient, high-quality global illumination with a small number of photons.
Our approach is particularly good at reconstructing diffuse-specular interactions like caustics, for which previous photon mapping methods require large photon sample counts (and path-tracing at reasonable sample counts can miss altogether).  
We focus on photon density estimation---a key component of all particle-based methods---and introduce a novel deep neural network that can estimate accurate photon density at any surface points in a scene given only sparsely distributed photons.

Previously, the most successful density estimation methods for photon mapping are kernel-based methods that use traditional kernel functions (like a uniform or cone kernel) to compute output radiance at a surface point as a weighted sum of nearby photons.
While previous methods have improved the kernels by controlling the kernel bandwidths or shapes \cite{kaplanyan2013adaptive,schjoth2008diffusion,kang2016adaptive}, traditional kernel functions still require a large enough count of photons located in a small enough bandwidth around every surface shading point, for which a very large number of photons need to be traced, to compute accurate photon density.
In contrast, {\em we propose to learn to predict a kernel function at each shading point to effectively aggregate nearby photon contributions}.
Our predicted kernels leverage data priors and are able to compute accurate photon density estimation for complex global illumination from photon counts that are an order of magnitude fewer than traditional methods.
%We demonstrate this in Fig.~\ref{fig:teaser}, where our result with 15k photons is better than different variations of path tracing and photon mapping \cite{jensen1996global} with significantly more samples and combined with image-space learning-based denoising.
%to progressive photon mapping with billions of photons \cite{hachisuka2008progressive}.

Our network considers local photons around a queried surface point within a predefined bandwidth as input.
Unlike traditional methods that often treat photons individually or leverage standard statistics to aggregate photons, {\em we leverage learned local photon statistics---encoded as a deep photon context vector inferred by the network---around a surface point for per photon kernel weight estimates}.
Specifically, the network first processes individual photons to extract per-photon features and aggregates them across photons using pooling operations to obtain a deep photon context feature that represents the local photon statistics.
The network processes the individual per-photon features concatenated with the local context to compute per-photon kernel weights, which are used to perform density estimation by a weighted sum.
We demonstrate that this approach of learning kernel prediction is more efficient than a baseline that directly estimates photon density from the aggregated deep context vector.

To train our network, we create diverse photon distributions by tracing photons in 500 procedurally generated scenes with complex shapes and materials.
We sample surface points on diffuse surfaces, which form a 512$\times$512 image (one pixel per point) in each scene, and we compute the ground truth photon density of each point using progressive photon mapping \cite{hachisuka2008progressive} with billions of photons.
%As a result, we have 70 images that involve more than one million surface points to train our network for density estimation.
Note that, our network focuses on local photon distribution properties of surface points.
Hence, every surface point in a scene is a training datum, allowing us to train a generalizable network without a lot of images. 

In Fig.~\ref{fig:teaser}, we demonstrate that, using only 15k photons, our method can synthesize high-quality images.
Conversely, variations of path tracing and photon mapping fail to do so; even when combined with advanced progressive and adaptive techniques, SPPM and APPM require significantly more samples (1.5M photons) to achieve comparable results.
%We demonstrate that our trained network can synthesize realistic images of novel scenes with only tens of thousands of photons; standard photon mapping is only able to approach our reconstruction quality with an order of magnitude more photons.
%is significantly fewer than billions of photons that are necessitated by stochastic progressive photon mapping.
%Our approach is the first that achieves sparse reconstruction in photon mapping and reduces the required numbers of samples by orders of magnitude. 
This makes our approach an important step towards making photon mapping computationally efficient.
Moreover, our experiments leverages an effective practical hybrid approach: using our method for reconstructing light-specular (LS) paths -- the light transport paths that interact with specular surfaces before arriving at light sources---and low sample-count path tracing with learning-based denoising for all other light transport paths.
This leverages the advantages of both MC denoising and our efficient photon density estimation technique.
%Finally, our network can be easily integrated in many existing photon mapping technique as a plug-and-play replacement for the kernel function in the photon density estimation.

\section{Related Work}
\label{sec:related}

\boldstart{Monte Carlo path integration.}
Kajiya \shortcite{kajiya1986rendering} introduced the rendering
equation and Monte Carlo (MC) path tracing.  Since then,
various methods for MC path integration have
been developed, including light tracing \cite{dutre1993monte},
bidirectional path tracing (BDPT)
\cite{lafortune1993bi,veach1995optimally}, and Metroplis light transport
(MLT) \cite{veach1997robust,pauly2000metropolis,cline2005energy}.
These methods are able to simulate complex light transport with
accurate global illumination in an unbiased way.  However, pure MC
based methods typically require a very large number of samples (traced
paths), especially for very low probability paths like the classical 
caustic or specular-diffuse-specular (SDS) paths.
%to reduce the variance in the rendered images for realistic
%results, and they may involve paths that have extremely low or even
%zero probability to be sampled, like the classical
%specular-diffuse-specular (SDS).  
We base our method on the
photon mapping technique, which is efficient for caustics and SDS,
%and SDS path computation, 
and we aim to achieve sparse reconstruction.
% with a small number of samples in  photon mapping.  
%And we aim to achieve sparse reconstruction for photon mapping, which
%uses only a small number of samples.

\boldstartspace{Monte Carlo denoising.}
While there is little progress in sparse reconstruction with low sample counts in photon mapping, 
many approaches have been proposed to achieve MC rendering with low sample counts. 
A recent survey of sparse sampling and reconstruction %in Monte Carlo rendering 
is presented by Zwicker et al. \shortcite{zwicker2015recent}.
MC denoising methods can be categorized into a-priori methods that rely on prior theoretical knowledge \cite{durand2005frequency,egan2009frequency,yan2015fast,wu2017multiple}, 
and a-posteriori methods that filter out the noise in rendered images with few assumptions about the image signal \cite{overbeck2009adaptive,rousselle2013robust,kalantari2015machine}.

Recently, deep learning techniques have been introduced to achieve MC denoising \cite{chaitanya2017interactive,bako2017kernel}, 
and many methods utilize kernel prediction\cite{bako2017kernel,vogels2018denoising,xu2019adversarial}.
Kalantari et al. \shortcite{kalantari2015machine} propose to predict the parameters of fixed filtering functions using fully-connected neural networks.
Bako et al. \shortcite{bako2017kernel} leverage deep convolution neural networks to predict kernels to linearly combine the original noisy radiances of neighboring pixels.
Gharbi et al. \shortcite{gharbi2019sample} make use of individual screen-space path samples and predict a kernel for each sample that splats the radiance contributions to its neighboring pixels.
In contrast, we apply deep learning in photon density estimation and leverage local photon statistics for density estimation from sparse photons. 
Our network considers individual scene-space photon samples around each shading point and predicts a kernel to gather per-photon contributions.
Our approach is the first that introduces deep learning in photon mapping and demonstrates learning-based kernel prediction in this context.

\boldstartspace{Photon density estimation.}
The rendering equation \cite{kajiya1986rendering,immel1986radiosity} can be approximated
by particle density estimation
\cite{shirley1995global,jensen1996global,walter1997global}.  Most
particle-based methods are based on the original photon mapping
framework \cite{jensen1996global}; it first traces rays from light
sources to distribute photons in a scene, and then gathers neighboring
photons at individual shading points to approximate radiance
estimates.  Photon mapping achieves low variance in the rendered
images and leads to blurred, less noticeable artifacts at the cost of
introducing bias in the estimates.  
%While photon mapping is a biased
%method, it's able to effectively handle challenging light transport
%settings including SDS paths.  In theory, 
Photon mapping is able to
consistently converge to the correct solution by increasing the number
of photons towards infinity and reducing the bandwidth towards zero.

\begin{figure*}[t]
    \includegraphics[width=\linewidth]{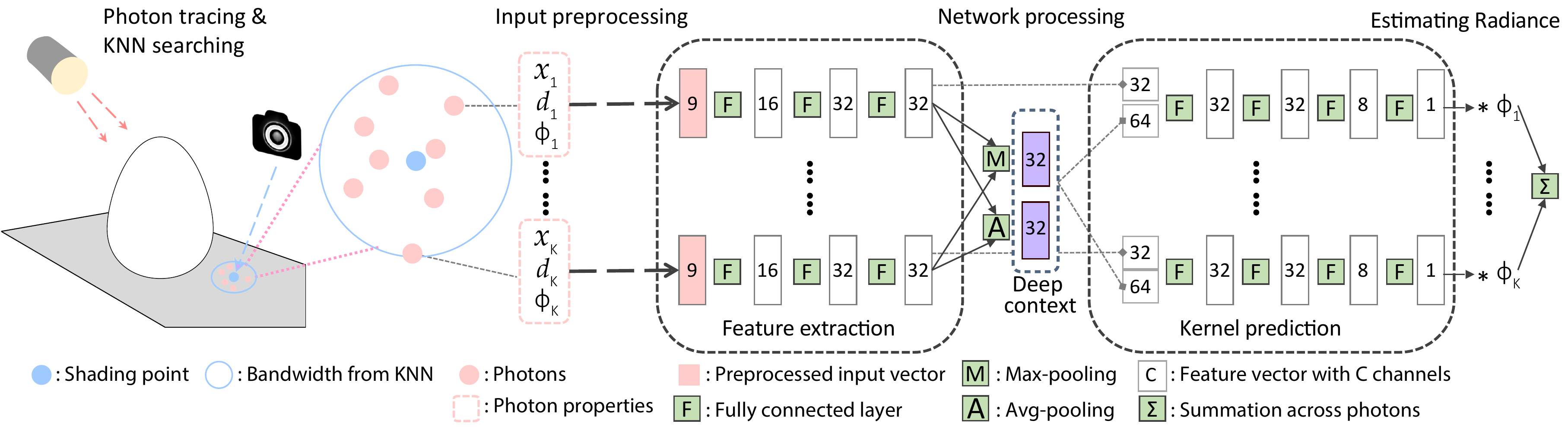}
    \caption{Overview of our deep photon density estimation network. Given a set of photons within the bandwidth of a shading point, we pre-process these photons' properties and input them to feature extractor MLPs that compute per-photon features. These are aggregated using max- and average-pooling to construct a deep context feature. The original per-photon features and the deep context are concatenated and processed by a kernel prediction MLP that predicts a kernel weight. Finally, these kernel weights are used to sum the photon contributions and produce the reflected radiance.}
    \label{fig:net}
\end{figure*}

Previous work has investigated 
progressive methods to overcome the memory bottleneck and enable arbitrarily large photon numbers \cite{hachisuka2008progressive,hachisuka2009stochastic,hachisuka2010progressive,knaus2011progressive},
bidirectional methods to improve rendering glossy objects \cite{vorba2011bidirectional}, 
adaptive methods to optimize photon tracing \cite{hachisuka2011robust},
and the combination of unbiased MC methods and photon mapping \cite{georgiev2011bidirectional,hachisuka2012path,georgiev2012light}.
Many relevant works have been presented to improve the kernel density estimation by utilizing standard statistics for adaptive kernel bandwidth \cite{jc95,kaplanyan2013adaptive,kang2016adaptive} or anisotropic kernel shapes \cite{schjoth2008diffusion}.
Other works leverage ray differentials \cite{schjoth2007photon}, blue noise distribution \cite{spencer2009into,spencer2013photon,spencer2013progressive}, and  Gaussian mixtures fitting \cite{jakob2011progressive} to improve the reconstruction. 
In contrast, we focus on accurately computing photon density with sparse photons, which hasn't been explored in previous work.
Essentially, we replace the traditional kernel density estimation with a novel learning based module, and keep the rest unchanged in the standard photon mapping framework.
This potentially enables the combination of our technique and previous photon mapping techniques that focus on other components in the framework.

%\section{Density estimation in photon mapping}
\section{Background: Density estimation}
\label{sec:density}
Photon mapping techniques compute reflected radiance via density estimation.
Kernel density estimation \cite{wand1994kernel} is the most widely used density estimation method in statistics, 
and has been widely applied in photon mapping.
Early works use the uniform kernel that treats nearby photons equally \cite{jensen1996global,hachisuka2008progressive};
subsequent works extend photon density estimation to support arbitrary smooth kernels \cite{hachisuka2010progressive,knaus2011progressive}.
In general, the reflected radiance at a shading location $\bm x$ is computed by:
\begin{align}
    L(\bm x,\bm \omega) \approx \frac{1}{N} \sum_{i=1}^{N} k_r(\bm x,\bm x_i) \tau_i,
    \label{eqn:kernel}
\end{align}
where $N$ is the total number of photon paths that are emitted in a scene, $\bm \omega$ is the reflected direction,
$\bm x_i$ is the location of a photon, $\tau_i$ is the photon contribution and $k_r$ represents the kernel function with bandwidth $r$.
In general, the photon contribution $\tau_i$ is the product of the BRDF and the photon energy. 
In this work, we only compute photon density on diffuse surfaces, as is done in many classical photon mapping methods.
In this case, the BRDF at a shading point is $\rho/\pi$, where $\rho$ is the albedo.
Correspondingly, $\tau_i=\phi \rho /\pi$, where $\phi$ represents the accumulated path contribution divided by the sampling probability, 
which can be also interpreted as the energy flux carried by the photon.
Therefore, $\bm \omega$ can be removed and $\rho$ can be taken out of the summation in Eqn.~\ref{eqn:kernel}.
We therefore consider the photon energy $\phi$ as the photon contribution in this work.

The kernel $k_r$ assigns linear weights to photons, which are used to linearly combine the contributions of photons in a local window with radius $r$.
Traditionally, $k_r$ is a uniform function ($1/(\pi r^2)$) or a function of the distance from the shading point to a photon ($\|\bm x-\bm x_i\|$).
Instead, we propose to leverage data priors to predict kernels to aggregate photon contributions.

\section{Learning to compute photon density}
\label{sec:learning}
In this section, we present our learning-based approach for photon density estimation.
Our approach is light-weight and focuses on density estimation only; 
we keep the main framework of standard photon mapping 
and upgrade the traditional, distance-based and photon-independent kernel functions ($k_r$ in Eqn.~\ref{eqn:kernel})
to novel, learned and local-context-aware kernel functions represented by a deep neural network (see Fig.~\ref{fig:net}).

In particular, given a shading point, our network considers its $\PhotonNN$ nearest neighbor photons, which adaptively selects the bandwidth $r$.
Multiple properties of individual photons are used as input for the network, 
including photon positions $\{\bm x_i\}_{i=1}^{\PhotonNN}$, 
photon directions $\{\bm d_i\}_{i=1}^{\PhotonNN}$ and photon contributions $\{\phi_i\}_{i=1}^{\PhotonNN}$.
We also supply the number of nearest photons $\PhotonNN$ to the network to let it better understand the local photon distribution.
Our network (denoted as $\Net$) regresses per-photon kernel weights to compute radiance estimates via a weighted sum similar to Eqn.~\ref{eqn:kernel}:
\begin{align}
    L(\bm x) \approx \frac{\rho}{N\pi r^2} \sum_{i=1}^{\PhotonNN} \Net_{r,i}(\bm x,\{\bm x_i\}, \{\bm d_i\}, \{\phi_i\}) \phi_i,
    \label{eqn:learning}
\end{align}
where $\Net_{r,i}$ represents the predicted kernel weight for photon $i$.
Note that, our network uses information about {\em all} photons in a local neighborhood for per-photon kernel prediction; 
it obtains deep photon statistics and 
associates per-photon information with statistical context to compute kernels for photon aggregation.

\subsection{Input pre-processing}
Photon distributions are highly diverse across shading points and across scenes, 
making it challenging to design a network that generalizes across different inputs.
Besides, deep neural networks are known to benefit from normalized input data to correlate values from different domains.
Therefore, we pre-process the input photon properties to allow for better generalizability and performance.

Since light intensities can have very high dynamic range (HDR), the photon contributions $\tau_i$ can vary widely in range,
which is highly challenging for a network to process.
We introduce a mapping function to pre-process the photon contributions, % the mapping function is given by 
\begin{align}
    t_a(u) = \frac{\log(u + a)- \log(a)}{\log(u+a) - \log(a) + 1},
    \label{eqn:mapping}
\end{align}
where $a=0.01$ is an additional parameter. 
Essentially, $t_a(u)$ maps HDR values $u$ from $[0,\infty]$ to $[0,1]$.
We further linearly map these values to $[-1, 1]$ and provide them as network input.
We observe that such a mapping process facilitates the network learning.

For photon positions $\bm x_i$ and directions $\bm d_i$, we first transform them into the local coordinate frame of the shading point; 
the coordinate frame is constructed using the position and normal of the shading point and two orthogonal directions that are randomly selected in the tangent plane.
This transforms the network inputs into a consistent coordinate system and improves generalizability.

%We also divide the number of nearest photons $\PhotonNN$ by $\PhotonNN_{\max} = 800$ that represents the highest number of nearest photons we consider;
%the scaled $\PhotonNN$ ($\in [0,1]$) is further linearly mapped into $[-1,1]$ before being input to the network.
The bandwidth $r$ of our learned kernel is determined by the distance of the $\PhotonNN^{\mathrm{th}}$ nearest photon.
This leads to a large range of bandwidth values given various photon distributions, 
which is highly challenging for a deep neural network to process.
Motivated by the bandwidth normalization used in traditional kernels \cite{wand1994kernel,shirley1995global}, 
we divide the photon positions in the local coordinates by the bandwidth $r$, 
and scale the final density estimates by $1/r^2$, which is shown in Eqn.~\ref{eqn:learning}.
This normalizes all input photon positions into a unit sphere and post-scales the computed photon density by the actual window area.
As a result, our network is invariant to the actual bandwidths, 
and effectively generalizes to different photon distributions and supports different numbers of \emph{total emitted photons} that will introduce different bandwidths for the same $\PhotonNN$.

Note that, different terms of our network input are all normalized into the range of $[-1,1]$, 
which enables our network to correlate and leverage different photon properties from various domains in an efficient way.
Our input pre-processing also makes our network translation-, rotation-, and scale- invariant to diverse photon distributions, 
leading to good generalization across different scenes and different numbers of emitted photons.

\subsection{Network architecture}
The inputs to our network are essentially a set of multi-feature 3D points in a unit sphere.
In the set, there is no meaningful inherent point ordering and the number of points ($\PhotonNN$) is not fixed.
We thus leverage PointNet \cite{qi2017pointnet} style neural networks with multi-layer perceptrons, 
which accept an arbitrary number of inputs and are invariant to permutations of inputs.
As shown in Fig.~\ref{fig:net}, our network consists of two sub-networks, a feature extractor and a kernel predictor; 
they are both fully connected neural networks and process each photon individually.

The feature extractor first processes each individual photon;
it considers the pre-processed photon properties (9 channels including positions, directions and contributions) as input, and extracts meaningful features using multilayer perceptrons.
Specifically, we use three fully connected layers in the feature extractor, and each layer is followed by a ReLu activation layer.
The feature extractor leverages linear and non-linear operations to transform the original input into a learned 32-channel feature vector.
These per-photon features are then aggregated across photons by max-pooling and average-pooling operations 
which output the deep photon context vector.  This vector represents the local photon statistics in a learned non-linearly transformed space.
The kernel predictor then leverages the across-photon context and the per-photon features 
to predict a single scalar that represents the kernel weight for each photon. 
These per-photon kernel weights are the final output of our network 
and will be used to linearly combine the original photon contributions as expressed in Eqn.~\ref{eqn:learning}.
The kernel predictor is also a three-layer fully connected neural network with ReLU as activation layers, 
which is similar to the feature extractor but with different channels at each layer.

Note that, unlike previous work that treats each photon independently, 
we propose to correlate per-photon information with local context information across photons.
Our feature extractor transforms photon properties into learned feature vectors,
which allows for collecting photon statistics in the learned neural feature space to obtain the photon context for the following kernel prediction.
Our whole network is very light-weight, and involves only six fully connected layers;
this ensures a highly efficient inference process.
We show that such a light-weight network is able to effectively reconstruct accurate photon density from sparse photons.

\subsection{Training details}

\boldstart{Data generation.}
Monte Carlo denoising usually requires a large number of images to train and is hard to generalize across different type of scenes.
Our method focuses on local photon distributions; 
in other words, to learn proper data priors, 
we desire the diversity of photon distributions in terms of individual shading points and not necessarily of the entire scenes.
This allows for good generalizability of our network with a relatively small number of training scenes,  
which can even be very different from our final testing scenes.
Inspired by \cite{xu2018deep,xu2019deep}, 
we procedurally create shapes from primitives with random sizes and random bump maps;
a set (randomly from 1 to 16) of such shapes are then placed in a box and distributed roughly as a grid.
We also place multiple area lights with random locations and rotations in the scene,
and randomly assign specular materials and diffuse materials to the scene objects.

\begin{figure}[h]
    \includegraphics[width=\linewidth]{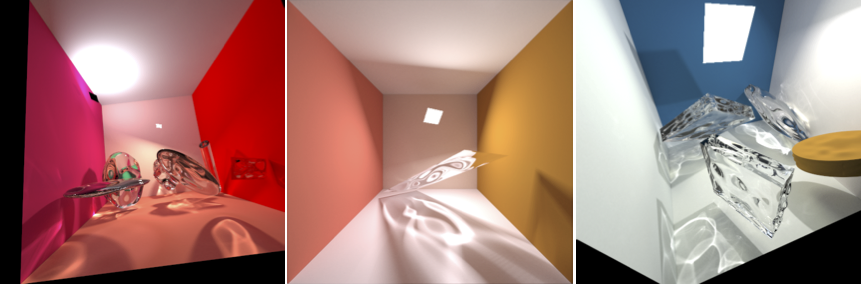}
    \caption{Examples of our procedurally generated training scenes.}
    \label{fig:scenes}
\end{figure}

A few examples of these scenes are shown in Fig~\ref{fig:scenes};
complex light transport effects with diverse photon distributions are simulated.
To sample shading points in each scene, we shoot rays from a camera through an image plane with $512\times 512$ pixels 
and select the first diffuse intersections as target shading points.
We trace photons from light sources and keep the ones that contribute to the indirect lighting.
Progressive photon mapping \cite{hachisuka2008progressive} is then applied
to compute ground truth photon densities for each point with a total number of about 1 billion photon paths. %\KS{Is this correct?}
For each scene, we store 10 million photon paths and a $512\times 512$ multi-channel image that contains the ground truth radiances 
and other necessary information (positions, normals and BRDFs) of shading points.
We create 500 scenes for training our neural networks and test our network on scenes that are significantly different from our training data (see Fig.~\ref{fig:teaser} and Fig.~\ref{fig:results}).

\boldstartspace{Loss function.}
We supervise our network with the ground truth radiance estimates.
The final radiances are in high dynamic range, which can easily make the training dominated by high-intensity values;
we therefore tone-map the radiance estimates using the $\mu$-law as in \cite{kalantari2017deep}. 
The mapping function $p_{\mu}(v)$ is given by:
\begin{align}
    p_{\mu}(v) = \frac{\log(1 + \mu v )}{\log(1+\mu)},
    \label{eqn:lossmapping}
\end{align} 
and we set $\mu = 5000$ following \cite{kalantari2017deep}.
%\KS{why not same function as Eqn. 3?}
We tone-map both our estimated radiance and the ground truth radiance, and we apply $L_2$ loss on the mapped values.

\boldstartspace{Training parameters.}
We randomly select $\PhotonNN$ from 100 to 800 and use from 0.3 million to 4 million photons to train our network,
which makes it generalize well to various bandwidths and photon counts.
We use Adam to train our network for 6000 epochs with an initial learning rate of $10^{-4}$ and a batch size of 2000 random shading points.

\Comment{
    given different number of emitted photons,

    As mentioned, we select $\PhotonNN$ nearest photons around each shading point, 
which determines t
This adaptive bandwidth selection has been widely used in traditional photon mapping to account for diverse photon density; 
however, it 
introduces highly different input scales for different numbers of emitted photons, 
}

\section{Experiments}
We now present a comprehensive evaluation of our method.
%by comparing our network with other network variants and showing results on novel scenes with complex caustic effects

\begin{figure}[h]
    \includegraphics[width=\linewidth]{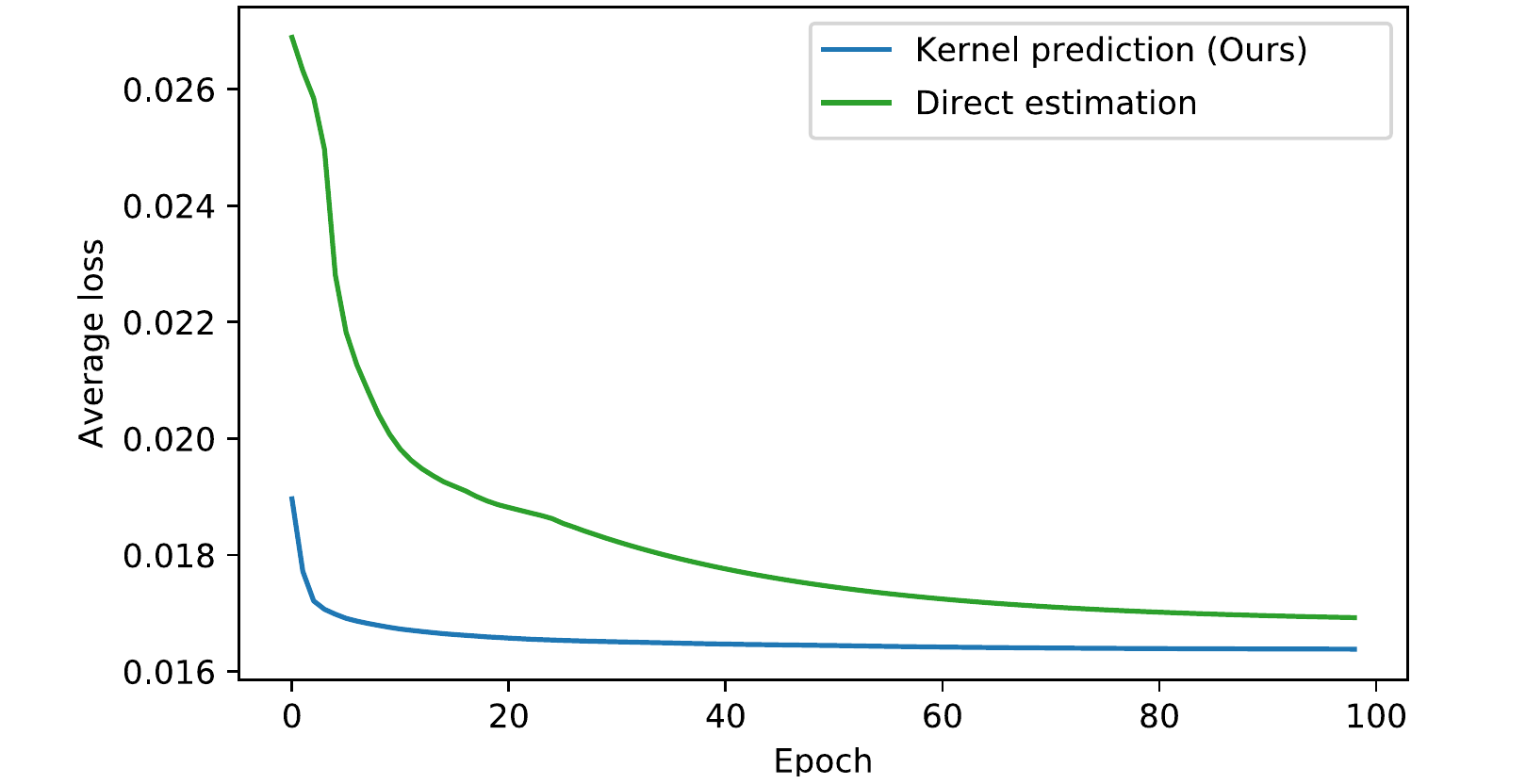}
    \caption{Optimization speed. We compare the optimization speed of our kernel-prediction network, and a baseline direct-estimation network.  Our network converges faster to the lower loss value.}
    \label{fig:opt}
\end{figure}

\boldstart{Ablation study.}
We first justify the choices of our network design.
In particular, 
we compare our network with a baseline network that estimates the final radiance without predicting kernels;
this comparison network has a similar network architecure but directly outputs the final irradiance from the across-photon deep context vector.
%We also compare against our network without using Eqn.~\ref{eqn:mapping} for normalization.
Figure~\ref{fig:opt} shows the training processes of these networks;
our network converges significantly faster than the baseline method.
This demonstrates the effectiveness of combining kernel density estimation and deep learning and is consistent with previous results on denoising for path tracing \cite{bako2017kernel,vogels2018denoising,gharbi2019sample}.

\boldstartspace{Evaluation scenes and photon generation.}
We evaluate our method on six challenging scenes (\Scenea, \Sceneb, {\Scenec}, \Scened, \Scenee, \Sceneg)
that involve complex caustics and other diffuse-specular interactions with LS paths.
In theory, LS paths can never be reconstructed by path tracing if we use a point light source; we therefore use area lights in the scenes to allow for reasonable comparisons with PT.
For each scene, we shoot photons for 0.1 second, which generates about 0.8M photon paths with at maximum five photons per path; we only keep those photons that involve light-specular paths in the scenes.
We denote the number of valid photons we consider as $M$, which is a number that is different from the total emitted photon paths $N$ in Eqn.~\ref{eqn:kernel}.
Because of various compositions of scenes, there are 15k (\Scenea), 85k (\Sceneb), 77k (\Scenec), 50k (\Scened), 100k (\Scened) and 125k (\Scened) valid photons that are used in the six scenes respectively.
We also evaluate with the number of photons that are traced in one second---corresponding to ten times the number of photons traced in 0.1 seconds---to justify the generalization of our network to different numbers of emitted photons, and compare with the other methods with photons that are traced in ten seconds to justify the quality of our sparse reconstruction.

\begin{figure}
    \includegraphics[width=\linewidth]{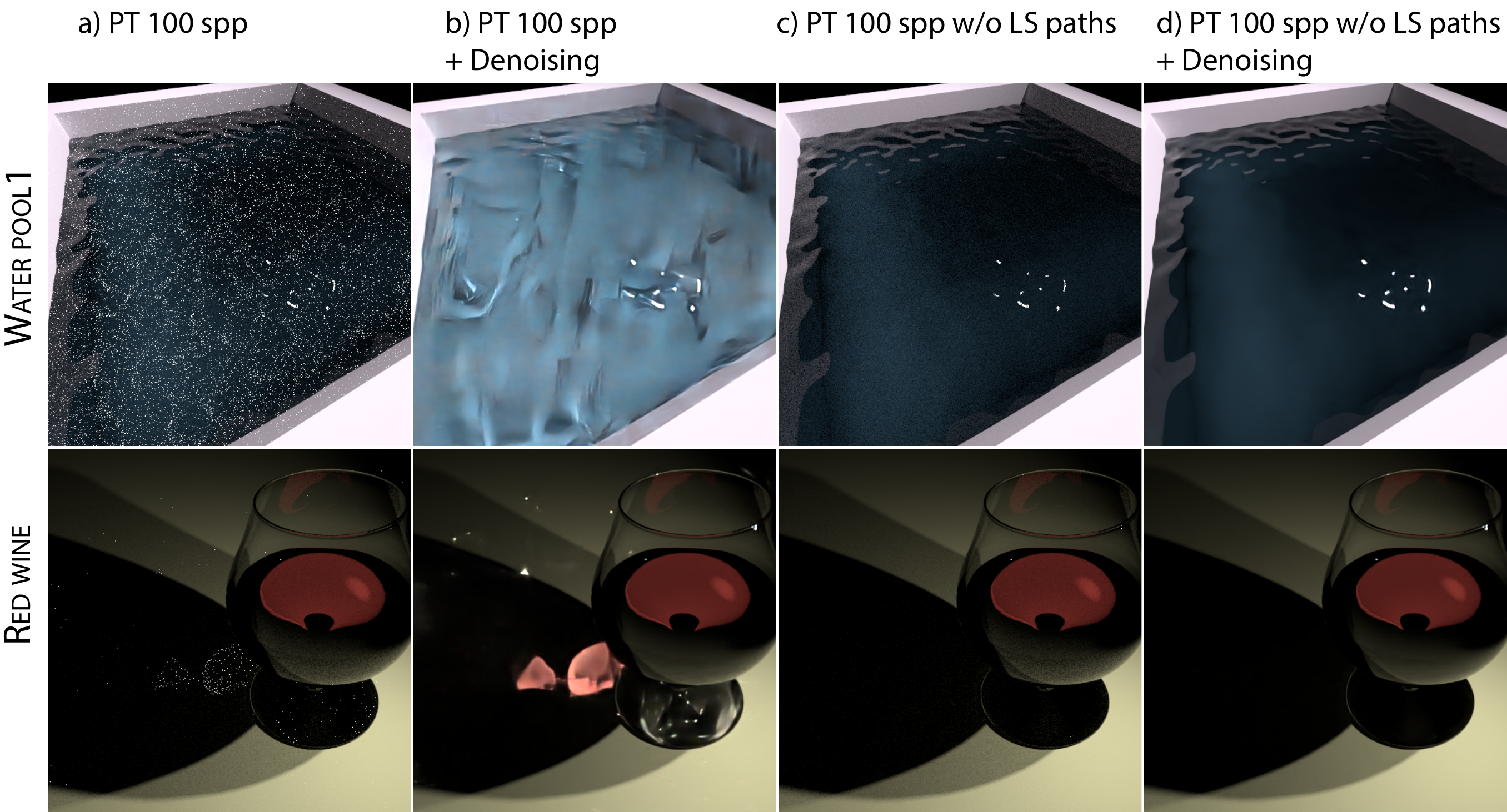}
    \caption{Path tracing and (PT) without light-specular paths (LS). We show PT and denoising results using 100 spp with and without light-specular paths.  The noise can be seen more clearly when 
    zooming into the electronic PDF.}
    \label{fig:pt}
    \vspace{-3mm}
\end{figure}

\newcolumntype{P}[1]{>{\centering\arraybackslash}p{#1}}
\newcolumntype{M}[1]{>{\centering\arraybackslash}m{#1}}

\newlength{\ourL}
\setlength{\ourL}{0.6in}
\begin{table*}[t]
	\begin{center}
		%\resizebox{\linewidth}{!}{
		
		\begin{tabular}{p{0.7in}|P{0.3in}P{\ourL}P{\ourL}P{0.4in}P{\ourL}P{\ourL}P{0.5in}P{0.4in}P{0.4in}| }  
			%\multicolumn{10}{|c|}{\Scenea}\\
			\hline  %\STAB{\rotatebox[origin=c]{90}{\Scenea}}     
			
			Scene & ($M$)  & Ours-50  & Ours-L-50  &  PM-50 
			& Ours-500 & Ours-L-500 &  PM-500 & PPM & APPM\\
			\hline \hline
			%\cline{2-10}
			\multirow{3}{*}{{\small \Scenea}}&(15k)    &   \cellcolor{blue!10} \textcolor{black}{0.013}       &    {0.006} \cellcolor{pink}             &  0.085
			&   \cellcolor{blue!10} {0.013}       &   \cellcolor{pink} 0.006             &  0.165       
			& 0.085  & \cellcolor{yellow!50}  0.080   \\
			\cline{2-10}
			&(150k)  &  0.012        &    {0.006}  \cellcolor{blue!10}          & \cellcolor{yellow!50} 0.036
			&  0.008        &    {0.004} \cellcolor{pink}           &  0.079        & 0.065 &   0.043  \\
			\cline{2-10}
			&(1.5M)  &  0.013        &     0.007           &  0.031
			&  \textcolor{black}{0.006}  \cellcolor{blue!10}      &    0.003  \cellcolor{pink}           & \cellcolor{yellow!50}  0.027        &  0.030  &  0.030  \\
			\hline
		\end{tabular}
	\Comment{
		\begin{tabular}{p{0.7in}|P{0.3in}P{\ourL}P{\ourL}P{0.4in}P{\ourL}P{\ourL}P{0.5in}P{0.4in}P{0.4in}| }  
			%\multicolumn{10}{|c|}{\Scenea}\\
			\hline  %\STAB{\rotatebox[origin=c]{90}{\Scenea}}     
			
			\multirow{3}{*}{{\small \Sceneaa}}&(15k)    &  0.015        &    0.005            &  0.104
			&  0.015        &    0.006            &  0.178       & 0.090   &  0.081   \\
			\cline{2-10}
			&(150k)  &   0.014       &     0.005           &  0.035
			&   0.010       &     0.005           &  0.097        & 0.064   &  0.044  \\
			\cline{2-10}
			&(1.5M)  &  0.016        &   0.006             &  0.028
			&  0.007        &   0.004             &   0.028       & 0.030    &  0.028  \\
			\hline
		\end{tabular}
	}
		\begin{tabular}{p{0.7in}|P{0.3in}P{\ourL}P{\ourL}P{0.4in}P{\ourL}P{\ourL}P{0.5in}P{0.4in}P{0.4in}| }  
			%\multicolumn{10}{|c|}{\Scenea}\\
			\hline  %\STAB{\rotatebox[origin=c]{90}{\Scenea}}     
			\multirow{3}{*}{{\small \Sceneb}}  &(85k)    &  0.052        &  \textcolor{black}{0.028}  \cellcolor{blue!10}            &  0.116
			&  0.044        &  {0.021}   \cellcolor{pink}            &  0.222       & 0.134 &  \cellcolor{yellow!50} 0.111   \\
			\cline{2-10}
			&(850k)  &   0.035       &      0.027          &  0.053
			&   \textcolor{black}{0.023}   \cellcolor{blue!10}    &      {0.014} \cellcolor{pink}          &  0.102        & 0.064  & \cellcolor{yellow!50}  0.047   \\
			\cline{2-10}
			&(8.5M)  &   0.032      &    0.030            &  0.045
			&   \textcolor{black}{0.014}   \cellcolor{blue!10}    &    {0.011}  \cellcolor{pink}          &  0.037        & 0.031 & \cellcolor{yellow!50}  0.026  \\
			\hline
		\end{tabular}
		
		\begin{tabular}{p{0.7in}|P{0.3in}P{\ourL}P{\ourL}P{0.4in}P{\ourL}P{\ourL}P{0.5in}P{0.4in}P{0.4in}|}  
			%\multicolumn{10}{|c|}{\Scenea}\\
			\hline  %\STAB{\rotatebox[origin=c]{90}{\Scenea}}     
			
			\multirow{3}{*}{{\small \Scenec}}&(77k)    &   0.042       &   \textcolor{black}{0.023}  \cellcolor{blue!10}           &  \cellcolor{yellow!50} 0.069
			&   \textcolor{black}{0.023}  \cellcolor{blue!10}     &   {0.008}  \cellcolor{pink}            &  0.153       & 0.137  &  0.143   \\
			\cline{2-10}
			&(770k)  &   0.041       &    0.024            &  0.046
			&   \textcolor{black}{0.011}   \cellcolor{blue!10}    &    {0.006}  \cellcolor{pink}           & \cellcolor{yellow!50} 0.042        &  0.050  &  0.049   \\
			\cline{2-10}
			&(7.7M)  &  0.045        &    0.020            &  0.066
			&   \textcolor{black}{0.012}   \cellcolor{blue!10}    &     {0.009}   \cellcolor{pink}         &  0.023       & 0.017   & \cellcolor{yellow!50} 0.014   \\
			\hline
		\end{tabular}
		
		\begin{tabular}{p{0.7in}|P{0.3in}P{\ourL}P{\ourL}P{0.4in}P{\ourL}P{\ourL}P{0.5in}P{0.4in}P{0.4in}|}  
			%\multicolumn{10}{|c|}{\Scenea}\\
			\hline  %\STAB{\rotatebox[origin=c]{90}{\Scenea}}     
			
			\multirow{3}{*}{{\small \Scened}}&(50k)    &  0.244        &   \textcolor{black}{0.174}  \cellcolor{blue!10}           &  0.281
			&  0.214        &   {0.146}  \cellcolor{pink}            &  0.323       & 0.327   & \cellcolor{yellow!50}  0.277  \\
			\cline{2-10}
			&(500k)  &   0.214       &   0.173             &  0.221
			&   \textcolor{black}{0.135}   \cellcolor{blue!10}    &   {0.115}  \cellcolor{pink}            &  0.244        & 0.249  &  \cellcolor{yellow!50} 0.193  \\
			\cline{2-10}
			&(5.0M)  &   0.237      &    0.186            &  0.259
			&   \textcolor{black}{0.107}   \cellcolor{blue!10}    &    {0.105}  \cellcolor{pink}           & \cellcolor{yellow!50}  0.124       & 0.206  &  0.125  \\
			\hline
		\end{tabular}
		
		\begin{tabular}{p{0.7in}|P{0.3in}P{\ourL}P{\ourL}P{0.4in}P{\ourL}P{\ourL}P{0.5in}P{0.4in}P{0.4in}|}  
			%\multicolumn{10}{|c|}{\Scenea}\\
			\hline  %\STAB{\rotatebox[origin=c]{90}{\Scenea}}     
			
			\multirow{3}{*}{{\small \Scenee}} &(102k)    &  0.178        &    {\textcolor{black}{0.125}}  \cellcolor{blue!10}          & 0.226 
			&   0.167       &     {0.095}  \cellcolor{pink}          &  0.260       & 0.262  & \cellcolor{yellow!50} 0.224   \\
			\cline{2-10}
			&(1.0M)  &  0.132        &    0.121            & \cellcolor{yellow!50} 0.147
			&  {\textcolor{black}{0.115}}  \cellcolor{blue!10}      &    {0.080}  \cellcolor{pink}           &  0.221        & 0.211  &   0.155  \\
			\cline{2-10}
			&(10.2M)  &   0.134       &   0.128             &  0.159
			&   {\textcolor{black}{0.066}}  \cellcolor{blue!10}     &      {0.061}  \cellcolor{pink}         &    0.102      & 0.163   & \cellcolor{yellow!50} 0.088  \\
			\hline
		\end{tabular}
		
		\begin{tabular}{p{0.7in}|P{0.3in}P{\ourL}P{\ourL}P{0.4in}P{\ourL}P{\ourL}P{0.5in}P{0.4in}P{0.4in}|}  
			%\multicolumn{10}{|c|}{\Scenea}\\
			\hline  %\STAB{\rotatebox[origin=c]{90}{\Scenea}}     
			
			\multirow{3}{*}{{\small \Sceneg}} &(125k)    &   0.066       &   0.054             & \cellcolor{yellow!50} 0.073
			&   {\textcolor{black}{0.052}}  \cellcolor{blue!10}     &    {0.043}  \cellcolor{pink}           &  0.089       &  0.126   &  0.102   \\
			\cline{2-10}
			&(1.2M)  &  0.056        &    0.054            &  0.061
			&  {\textcolor{black}{0.034}}  \cellcolor{blue!10}      &    {0.033} \cellcolor{pink}            & \cellcolor{yellow!50} 0.044        & 0.083    &   0.054  \\
			\cline{2-10}
			&(12.5M)  &   0.059       &     0.059           &  0.078
			&    {\textcolor{black}{0.028}}  \cellcolor{blue!10}    &     {0.027}  \cellcolor{pink}          & \cellcolor{yellow!50}  0.031       & 0.059  &  0.035   \\
			\hline
		\end{tabular}
	\Comment{
		\begin{tabular}{ p{0.7in}|P{0.3in}P{\ourL}P{\ourL}P{0.4in}P{\ourL}P{\ourL}P{0.5in}P{0.4in}P{0.4in}|}  
			%\multicolumn{10}{|c|}{\Scenea}\\
			\hline  %\STAB{\rotatebox[origin=c]{90}{\Scenea}}     
			
			\multirow{4}{*}{{\small \Scenef}} &(66k)    &  0.072        &    0.052            &  0.138
			&  0.047        &    0.029            &  0.234       & 0.116  &  0.114   \\
			\cline{2-10}
			&(660k)  &  0.051        &    0.048            &  0.068
			&  0.025        &    0.017            &  0.112        & 0.055  &  0.049   \\
			\cline{2-10}
			&(6.6M)  &  0.052        &   0.051             &  0.073
			&   0.016       &    0.015            &  0.036        & 0.024   &  0.024   \\
			\hline
		\end{tabular}
	}
		%}
		\caption{Quantitative RMSE evaluation. We test our networks trained with different $\PhotonNN$ ($\PhotonNN=50$ and $500$, denoted with Ours-$\PhotonNN$) on six novel scenes with different numbers of valid photons ($M$). We also test a variant of our network architecture with enlarged four times capacity (Ours-Large) using the same $\PhotonNN$.
			We compare RMSE against standard photon mapping (PM) \cite{jensen1996global} under the same conditions, and also progressive PM (PPM) \cite{hachisuka2008progressive} and adaptive PPM (APPM) \cite{kaplanyan2013adaptive}. We highlight the best and the second best results in red and blue for each row; note that, all of them are our results. We also highlight the best result of the comparison methods in yellow, which is often worse than any of our network settings. \Comment{Photon mapping's performance varies widely based on the settings but the best results (shown in \blue{blue}) are always with high photon count and $K=50$ nearest neighbors. On the other hand, our results are more consistent with different settings, demonstrating our networks' ability to adapt to different numbers of traced photons. Moreover, our worst results (shown in \red{red}) are usually with low photon counts and are similar to, if not better than the best photon mapping results (with 10x more photons). Meanwhile, our best results (shown in \blue{blue}) are at higher photon counts and have substantially lower errors than photon mapping with the same photon counts.}}
		\label{tab:rmses}
	\end{center}
\end{table*}

\boldstartspace{Combining MC denoising and deep photon mapping.}
We evaluate our deep photon density estimation by combining our method with MC denoising. 
Specifically, we apply our learning-based density estimation to only compute the challenging light transport effects
which involve LS paths that are extremely hard to trace in PT and likely to introduce caustics.
In addition, we use path tracing with relatively low sample counts to compute the remaining light transport paths, 
and use modern learning-based denoising---the Optix built-in denoiser based on \cite{chaitanya2017interactive}---to remove the MC noise.

\begin{figure}[t]
    \includegraphics[width=\linewidth]{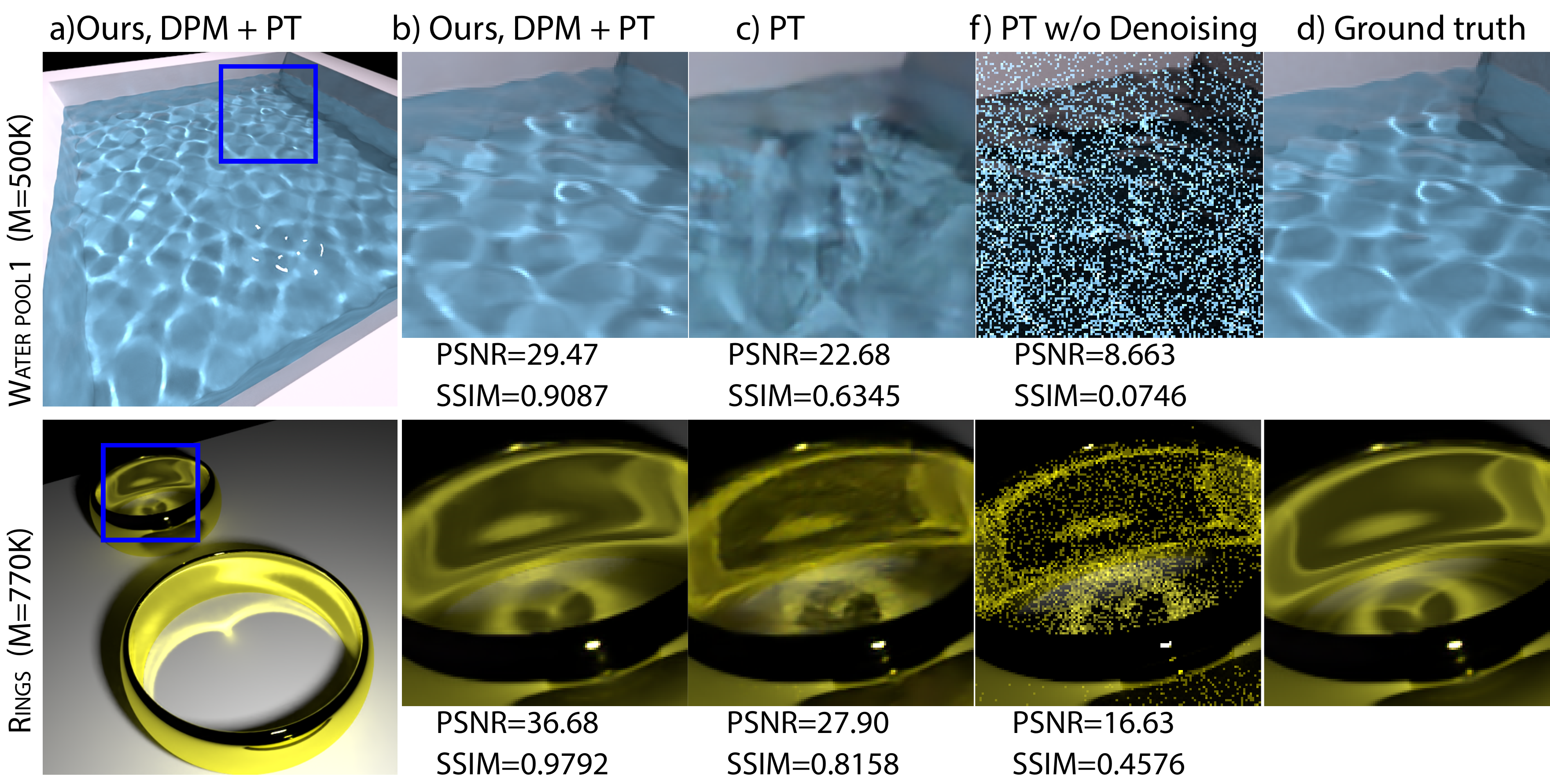}
    \caption{We show our final results in full images (a). 
    Our final results are computed by combining our deep photon mapping results and path tracing with denoising. We compare against pure path tracing using 1000 spp with (c) and without(d) denoising on insets.
    Obviously, path tracing alone even with 1000 spp cannot handle the LS paths.}
    \label{fig:vsPT}
    \vspace{-3mm}
\end{figure}

By removing LS paths in PT, we also make PT and MC denoising much easier. 
As shown in Fig.~\ref{fig:pt}, 
PT without LS paths can be effectively denoised using modern learning-based denoising techniques with 100 spp, whereas full PT with LS paths introduces extensive noise with the same 100 spp, causing denoising to fail completely.
In fact, the standard PT plus denoising pipeline is not able to recover the complex light transport effects with even 1000 spp (see Figs.~\ref{fig:teaser},\ref{fig:vsPT}).
In contrast, we demonstrate a practical way of combining our efficient deep photon mapping with MC denoising for photorealistic image synthesis, in which we leverage the benefits of low-sample reconstruction in both scene-space particle density estimation and screen-space MC integration.

\begin{figure*}
    \includegraphics[width=\linewidth]{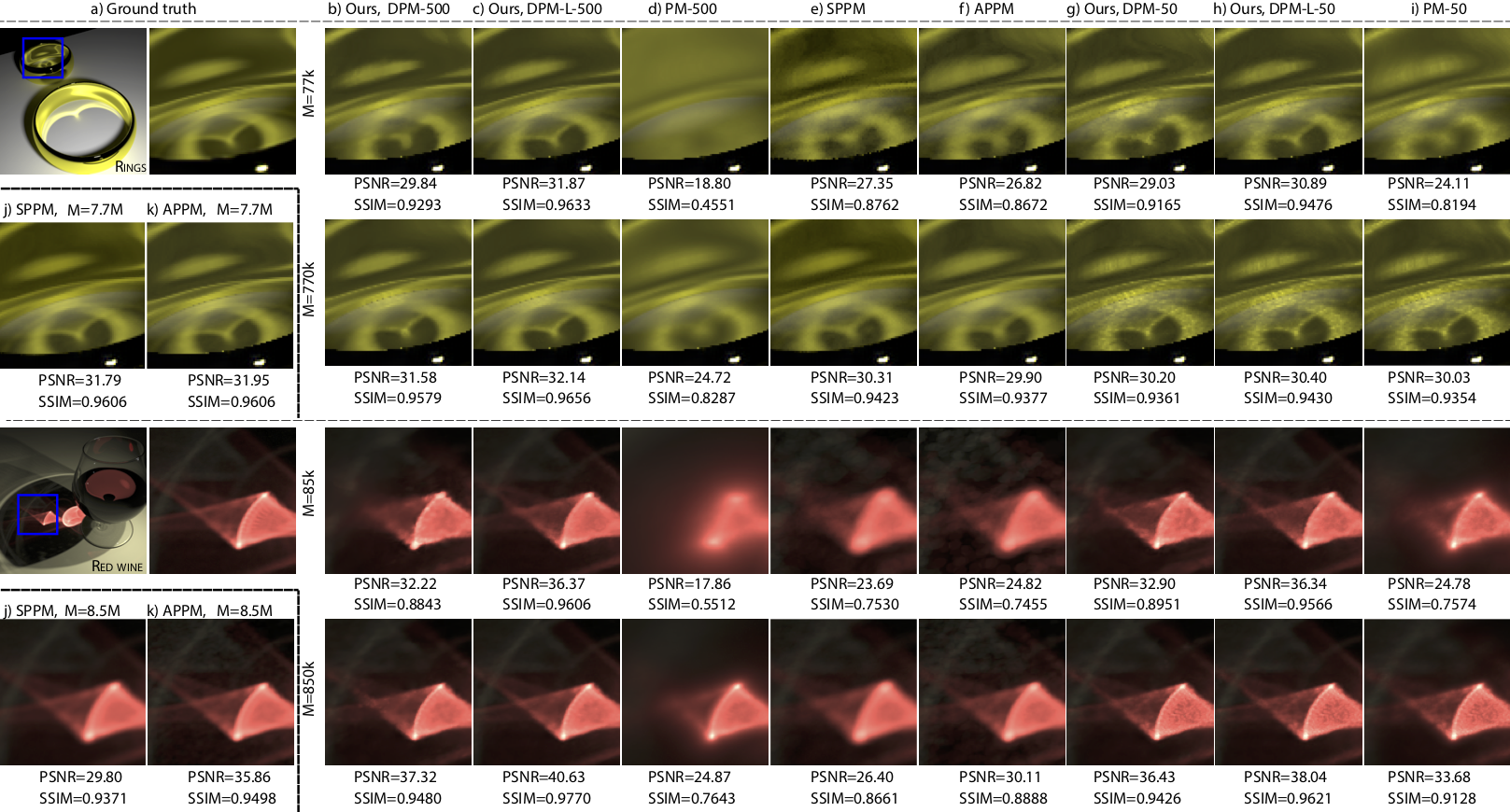}
    \caption{ We show results of our method with different numbers of input photons ($\PhotonNN$). We compare against PM, SPPM and APPM with the same number of total photons ($M$) on insets marked in the left-top ground truth image. We also show the results of APPM and PPM with ten times the largest number of photons our method uses (j, k). The PSNRs and SSIMs of the insets are shown correspondingly.}
    \label{fig:resultsNK}
\end{figure*}

\begin{figure*}[t]
    \includegraphics[width=\linewidth]{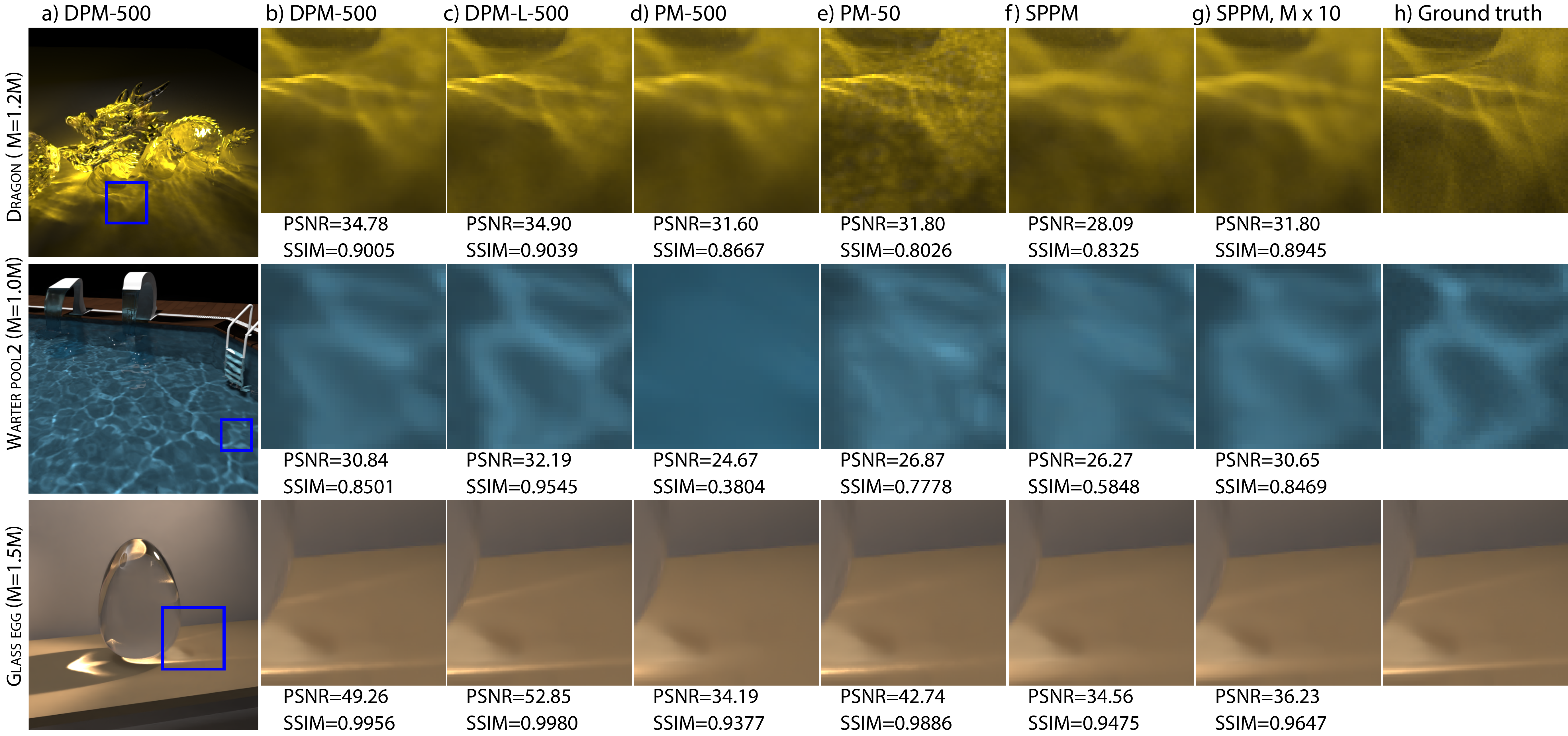}
    \caption{We show our results on full images (a). We compare against PM with the same input photons (d) and SPPM with the same (f) and ten times (g) the total photon counts on insets. PSNRs and SSIMs are also calculated for all insets and listed below.}
    \label{fig:results}
    %\vspace{-3mm}
\end{figure*}

\Comment{
We show our results on full images (a), and compare against PM with the same photons (c) and SPPM with ten times the photon counts (d )and the ground truth (e) with insets. The images purely generated with photon mapping techniques for LS paths are shown in (f)-(i) respectively. We combine these images with PT denoising using 300 spp for \Scenea~and 100 spp for the other three scenes. PSNRs and SSIMs are also calculated for all insets and listed below.}

\begin{table*}[t]
	\begin{center}
		%\resizebox{\linewidth}{!}{
        \begin{tabular}{ |c c c | c c c c |}  
        %\multicolumn{10}{|c|}{\Scenea}\\
        \hline  %\STAB{\rotatebox[origin=c]{90}{\Scenea}}     
          
               Photon tracing& Photon gathering & Number of photons & DPM-50  & DPM-L-50  & DPM-500 & DPM-L-500 \\
                    \hline
                  0.1s       &      0.12s$\sim$ 0.5s & 15k$\sim$ 125k  &  0.3s
                  &    1.0s      &     3.0s           &   10.0s
                  \\
                  \hline
                  1.0s       &      1.2s$\sim$ 5.0s & 150k$\sim$ 1.2M &  0.3s
                  &    1.0s      &     3.0s           &   10.0s
                  \\
                  \hline
                  10.0s       &      12s$\sim$ 50s & 1.5M$\sim$ 12M  &  0.3s
                  &    1.0s      &     3.0s           &   10.0s
                  \\
                  \hline
        \end{tabular}  
        \caption{Timing. We show the corresponding running time in seconds for each photon mapping component. Our experiments are run with photons that are traced within 0.1s, 1.0s and 10.0s in each scene. We list the corresponding gathering time to find the neighboring photons for about $512\times 512$ surface shading points.
        The numbers of total photons are also shown, corresponding to the $M$ in Tab.~\ref{tab:rmses}. 
        We list the network inference time for $512\times 512$ surface shading points for our regular network (DPM) and a large network (DPM-L) with $\PhotonNN=50$ and 500. Note that, the network inference time is determined by its capacity and $\PhotonNN$, and is independent of the number of total photons in the scene. }
		\label{tab:timing}
		\vspace{-3mm}
	\end{center}
\end{table*}

\begin{table*}[t]
	\begin{center}
		%\resizebox{\linewidth}{!}{
        \begin{tabular}{ p{0.7in}| c c c c c c c| }  
        %\multicolumn{10}{|c|}{\Scenea}\\
        \hline  %\STAB{\rotatebox[origin=c]{90}{\Scenea}}     
         \multirow{2}{*}{{\small Mean DSSIM}} 
           & Ours-50  & Ours-Large-50  &  PM-50 
                  & Ours-500 & Ours-Large-500 &  PM-500\\
                    \cline{2-7}
                    &   0.0346       &      0.0342          &  0.0337
                  &    0.0281      &     0.0277           &   0.0260  \\
                  \cline{2-7}

        \hline
        \end{tabular}  
        \caption{Temporal stability. We show the mean DSSIM between pairs of adjacent frames over a sequence of 30 rendered frames. Results have been averaged over all the different scenes and amount of photons.}
		\label{tab:temporal}
		\vspace{-1mm}
	\end{center}
\end{table*}

%Note that, since our networks always deal with photons normalized in a unit sphere as input, each network is well generalized across different radii with different number of total photons.

%\paragraph{Combining MC denoising and deep photon mapping.}
%Path tracing (PT) with MC denoising has been widely used in fast rendering with global illumination;
%however, complex effects with light-specular (LS) paths, like caustics are very difficult for this denoising solution to simulate, 
%which typically requires the involvement of photon mapping for efficient computation.

\boldstartspace{Parameters of our network and comparison methods.}
We observe that it is very hard for a single network to generalize across different numbers of input photons ($\PhotonNN$).
We thus use a fixed $\PhotonNN$ when training per network, and specifically we train two networks with $\PhotonNN=50$ and $\PhotonNN=500$ for the evaluation.
We also compare with a variant of our network that has four times the channels at each layer in our network architecture to evaluate if larger network capacity leads to higher performance. 
This large network generally leads to better performance (see Tab.~\ref{tab:rmses}), but it requires about three times longer inference time (see Tab.~\ref{tab:timing}); please see the following parts in the section for more discussion about quality and performance.
In the experiments, we use DPM (deep photon mapping) to denote the network with regular capacity and DPM-L (or Ours-L) to denote the one with larger capacity.

In all experiments, we compare with the classical photon mapping (PM) with the same k-NN photons as inputs.
We also compare with various progressive methods that are designed to progressively reduce the bandwidth with large photon counts.
In particular, for density estimation at fixed surface points, we compare with progressive photon mapping (PPM) \cite{hachisuka2008progressive}. 
Given a certain number of input photons, 
the quality of PPM is influenced by the initial radius and the number of photons per iteration.
To make a fair comparison, we compare 30 different variants (10 radii and 3 photon counts per iteration) of the two parameters and choose the best settings (with lowest RMSEs) for each scene.
We also compare with adaptive progressive photon mapping \cite{kaplanyan2013adaptive} similarly using the best radius and number of photons per iteration from 30 different variants of parameters.
For visual comparisons, we compare with stochastic PPM (SPPM) \cite{hachisuka2009stochastic}, when there are transparent surfaces in a scene which require sampling multiple surface points per pixel.

%\boldstartspace{Evaluation on different input settings.}
\boldstartspace{Quantitative and qualitative evaluation.}
We now evaluate our method quantitatively and qualitatively with different numbers of
photons counts ($M$) and different variations of training parameters (input photon number $\PhotonNN$ and capacity).
Table ~\ref{tab:rmses} shows quantitative RMSE evaluation of photon density estimation on the six testing scenes; the numbers are averaged across about 260k surface shading points sampled by tracing rays from a camera and selecting the first diffuse hit points in the scenes.
Note that, across all these different scenes with different photon counts, 
our method with $\PhotonNN=500$ performs consistently better than all the comparison PM methods, including standard PM  \cite{jensen1996global}, PPM \cite{hachisuka2008progressive} and APPM \cite{kaplanyan2013adaptive}, with the same number of total photons.
Most of our results are better than PM's and PPM's results with ten times the photon counts as ours.
APPM leverages traditional statistical information of local photons to improve the density estimation of PPM, which is able to achieve fairly good results; however, it requires the number of photons to be large enough to obtain good statistics.
In contrast, our method leverages learned statistics in the network, which achieves significantly better results than APPM with the same number of photons;
ours is actually comparable to the APPM that uses ten times the total number of photons.
Note that, the APPM and PPM results are selected from the results of tens of APPM and PPM variants with different hyper-parameters for their best performance; yet, our method still outperforms the best of these variants.

To visually illustrate the numbers in Tab.~\ref{tab:rmses}, we demonstrate all the rendering results of \Scenec~and~\Sceneb~with the first two rows (first two $M$) in Fig.~\ref{fig:resultsNK}; we also show the visual results of APPM and PPM with larger $M$ in Fig.~\ref{fig:resultsNK}.j,~k.
Additionally, we show results of three testing scenes in Fig.~\ref{fig:results}, where we compare our our DPM-500 with PM and SPPM.
In Fig.~\ref{fig:teaser}, we show the result of our DPM-50 and compare with PT, PM, SPPM and APPM.
In general, our method with $\PhotonNN=500$ outperforms the comparison methods with the same number of photons qualitatively and quantitatively. 
And our results are comparable to (if not better than) the comparison methods that use ten times the number of photons in the scene. 
While the larger network with $\PhotonNN=500$ (Ours-L-500) performs better than the regular network, the larger one also requires longer inference time (see Tab.~\ref{tab:timing}).
Therefore, our regular network with $\PhotonNN=500$ is generally the best choice for most cases, which stably achieves high-quality results.
However, when timing is not a critical issue, the large network will be a better choice for higher accuracy.

In most cases, our network is in favor of more nearest neighbor photons ($\PhotonNN$) as input to the network; the Ours-500 results are usually better than the Ours-50 ones.
Essentially, a larger $\PhotonNN$ allows for better local deep statistics in the deep context feature, which enables better kernel predictions.
Note that, this is not the case for standard PM using the same nearest neighbor strategy for bandwidth selection.
Photon mapping either introduces obvious non-smooth artifacts with a small bandwidth (Fig.~\ref{fig:resultsNK}.i) or outputs over-smooth results without details with a large bandwidth (Fig.~\ref{fig:resultsNK}.d).
APPM tends to resolve this issue by wisely reducing the bandwidth according to the photon statistics.
In contrast, our method achieves significantly better results than APPM when there are only sparsely distributed photons. 
Our method is able to leverage a relatively large bandwidth without introducing any obvious over-smoothing issues.
This is thanks to our learning based context-aware kernel prediction approach.
In particular, our approach allows for every single photon to leverage across-photon information in the learned deep context feature to tell if it is an outlier or an important contributing element to the shading point's reflected radiance; a corresponding kernel weight is assigned to each photon based on the decision made by data priors in the network.
Therefore, our method is able to effectively utilize the sparse photons in a large area to generate photorealistic images that are of high smoothness and have many details.

\boldstartspace{Timing.}
We use Optix to trace photons and do path tracing for all the results.
All experiments are run on one NVIDIA 1080 Ti GPU.
Path tracing runs at about 50 spp per second in all six scenes with an image resolution of 512$\times 512$.
It takes about 0.1, 1.0 and 10 seconds to emit photons.
We show the corresponding photon gathering time and network inference time for $512\times 512$ surface shading points in Tab.~\ref{tab:timing}.
In particular, we build Kd-Trees to do the neighboring search at each shading point and all methods take similar time to gather neighboring photons.
Note that the running time of our network is linear with the number of input photons $\PhotonNN$; it is also determined by the number of shading points that are required to be computed, and the listed timing corresponds to $512\times 512$ shading points.
The total running time for our method is the summation of the photon tracing, gathering and the network inference time; the total running time for the other methods is the summation of tracing and gathering.
Note that, across all the experiments (Tab.~\ref{tab:rmses}, Fig.~\ref{fig:resultsNK}, Fig.~\ref{fig:results}), our results of DPM-500 with photons traced in 1 second are comparable to the best results of comparison methods with photons traced in 10 seconds;
however, to achieve the comparable results, our DPM-500 takes about 5.2s$\sim$ 9s total time, whereas the comparison methods require 22s$\sim$ 60.0s total time to compute the same number of shading points.
Our method takes significantly shorter time to achieve the comparable quality.

\boldstartspace{Network capacity and $\PhotonNN$.}
While our network is mainly trained with relatively sparse photons (small $M$),
our network with $\PhotonNN$=500 overall generalizes well across different numbers of total photons ($M$) and, in most cases, achieves better performance when $M$ increases.
However, for $\PhotonNN$=50, there is too little information for the network to leverage and higher performance is often not ensured with a larger $M$.
Nonetheless, our network with $\PhotonNN$=50 still works well and performs better than the comparison methods when there are tens of thousands of photons.
We also observe that a larger network (Ours-L) with larger capacity leads to clearly better results than our regular network. Of course, a larger network requires higher computational cost or longer inference time as shown in Tab.~\ref{tab:timing}. Yet, the larger network with $\PhotonNN=50$ can already often achieve reasonably good results, which takes shorter running time than $\PhotonNN=500$. We leave the exploration of more variants of the network capacity and $\PhotonNN$ as future work.

\boldstartspace{Temporal consistency.}
Since our method deals with shading points in 3D space and is independent of view directions, we have observed that it has good across-frame consistency when changing the view in a scene with a fixed set of photons. 
%Similar to previous approaches for measuring temporal stability \cite{vogels2018denoising}, we move the camera and calculate the mean DSSIM between pairs of consecutive frames over a sequence of 30 frames. 
We follow \cite{vogels2018denoising} and use the mean DSSIM between consecutive frames to evaluate the temporal consistency when moving the camera.
Results in Table \ref{tab:temporal} show comparable temporal stability between our results and standard PM outputs. 
We leave the extensions of our network to recurrent architectures and general temporal consistency with other dynamic components in a scene as future work.

\boldstartspace{Progressive density estimation.}
Our current framework requires a fixed number of input photons for each trained network.  Progressive photon mapping accepts different numbers of photons per iteration with reduced bandwidth. 
Nonetheless, we have demonstrated that our network architecture supports accurate photon density estimation
with various fixed photon numbers. 
In other words, a progressive method can potentially be achieved by training a sequence of networks with different numbers of inputs.
A universal network for any given number of input photons may require introducing recurrent networks in the framework, which is an interesting direction of future work. 

%\paragraph{Network settings.}
%Note that, since our network deals with individual photons, the running time is linear with $\PhotonNN$. 
%In particular, it takes about $0.2$s with $\PhotonNN=50$ and $2$s with $\PhotonNN=500$ for our network to compute a $512 \times 512$ image on an NVIDIA 1080TI GPU. 

%\paragraph{Network settings.}
%Note that, since our network deals with individual photons, the running time is linear with $\PhotonNN$. 
%In particular, it takes about $0.2$s with $\PhotonNN=50$ and $2$s with $\PhotonNN=500$ for our network to compute a $512 \times 512$ image on an NVIDIA 1080TI GPU. 

%where every photon can its kernel weight and tell if it is an outlier for the shading point by leveraging across-photon information in the deep context feature. 

%which is not a very bad choice for a scene whose LS paths are mostly focused on small regions with thin-structure caustics like   

%We also demonstrate the corresponding renderings of the the scene \Scenea and \Scenec with the different input settings in Fig.~\ref{fig:resultsNK}.

\section{Conclusions and Future Work}
In this paper, we have presented the first deep learning-based method for density estimation in particle-based rendering. We introduce a deep neural network that learns a kernel function to aggregate photons at each shading point and renders accurate caustics with significantly fewer photons than previous approaches, with minimal overhead. Learning-based MC denoising has significantly improved path tracing results and our work extends these benefits to the popular photon mapping method.  

%In the future, we would like to connect the particle-based denoising approach for photon mapping with the more image-space approaches in use for path tracing; we have already shown examples in our results of combining the techniques. % essentially in terms of separate passes. 
Our method could be improved in the future with more advanced machine learning approaches, perhaps based on generative adversarial networks (GANs), just as has been done with path tracing \cite{xu2019adversarial}.  More broadly, we believe this paper points towards denoisers specialized to many other approaches for realistic image synthesis such as Metropolis Light Transport and Vertex Connection and Merging.

\bibliographystyle{eg-alpha-doi}  
\bibliography{reference}        

\newcommand{\etalchar}[1]{$^{#1}$}
\begin{thebibliography}{\uppercase{WYKR17}}

\bibitem[BVM{\etalchar{*}}17]{bako2017kernel}
\textsc{Bako S., Vogels T., McWilliams B., Meyer M., Nov{\'a}k J., Harvill A.,
  Sen P., Derose T., Rousselle F.}:
\newblock Kernel-predicting convolutional networks for denoising monte carlo
  renderings.
\newblock \emph{ACM Transactions on Graphics (TOG) 36}, 4 (2017), 97.

\bibitem[CKS{\etalchar{*}}17]{chaitanya2017interactive}
\textsc{Chaitanya C. R.~A., Kaplanyan A.~S., Schied C., Salvi M., Lefohn A.,
  Nowrouzezahrai D., Aila T.}:
\newblock Interactive reconstruction of monte carlo image sequences using a
  recurrent denoising autoencoder.
\newblock \emph{ACM Transactions on Graphics (TOG) 36}, 4 (2017), 98.

\bibitem[CTE05]{cline2005energy}
\textsc{Cline D., Talbot J., Egbert P.}:
\newblock Energy redistribution path tracing.
\newblock In \emph{ACM Transactions on Graphics (TOG)} (2005), vol.~24, ACM,
  pp.~1186--1195.

\bibitem[DHS{\etalchar{*}}05]{durand2005frequency}
\textsc{Durand F., Holzschuch N., Soler C., Chan E., Sillion F.~X.}:
\newblock A frequency analysis of light transport.
\newblock \emph{ACM Transactions on Graphics (TOG) 24}, 3 (2005), 1115--1126.

\bibitem[DLW93]{dutre1993monte}
\textsc{Dutr{\'e} P., Lafortune E.~P., Willems Y.~D.}:
\newblock Monte carlo light tracing with direct computation of pixel
  intensities.

\bibitem[ETH{\etalchar{*}}09]{egan2009frequency}
\textsc{Egan K., Tseng Y.-T., Holzschuch N., Durand F., Ramamoorthi R.}:
\newblock Frequency analysis and sheared reconstruction for rendering motion
  blur.
\newblock In \emph{ACM Transactions on Graphics (TOG)} (2009), vol.~28, ACM,
  p.~93.

\bibitem[GKDS12]{georgiev2012light}
\textsc{Georgiev I., Kriv{\'a}nek J., Davidovic T., Slusallek P.}:
\newblock Light transport simulation with vertex connection and merging.
\newblock \emph{ACM Trans. Graph. 31}, 6 (2012), 192--1.

\bibitem[GKS11]{georgiev2011bidirectional}
\textsc{Georgiev I., K{\v{r}}iv{\'a}nek J., Slusallek P.}:
\newblock Bidirectional light transport with vertex merging.
\newblock In \emph{SIGGRAPH Asia 2011 Sketches} (2011), ACM, p.~27.

\bibitem[GLA{\etalchar{*}}19]{gharbi2019sample}
\textsc{Gharbi M., Li T.-M., Aittala M., Lehtinen J., Durand F.}:
\newblock Sample-based monte carlo denoising using a kernel-splatting network.
\newblock \emph{ACM Transactions on Graphics (TOG) 38}, 4 (2019), 1--12.

\bibitem[HJ09]{hachisuka2009stochastic}
\textsc{Hachisuka T., Jensen H.~W.}:
\newblock Stochastic progressive photon mapping.
\newblock In \emph{ACM Transactions on Graphics (TOG)} (2009), vol.~28, ACM,
  p.~141.

\bibitem[HJ11]{hachisuka2011robust}
\textsc{Hachisuka T., Jensen H.~W.}:
\newblock Robust adaptive photon tracing using photon path visibility.
\newblock \emph{ACM Transactions on Graphics (TOG) 30}, 5 (2011), 114.

\bibitem[HJJ10]{hachisuka2010progressive}
\textsc{Hachisuka T., Jarosz W., Jensen H.~W.}:
\newblock A progressive error estimation framework for photon density
  estimation.
\newblock In \emph{ACM Transactions on Graphics (TOG)} (2010), vol.~29, ACM,
  p.~144.

\bibitem[HOJ08]{hachisuka2008progressive}
\textsc{Hachisuka T., Ogaki S., Jensen H.~W.}:
\newblock Progressive photon mapping.
\newblock In \emph{ACM Transactions on Graphics (TOG)} (2008), vol.~27, ACM,
  p.~130.

\bibitem[HPJ12]{hachisuka2012path}
\textsc{Hachisuka T., Pantaleoni J., Jensen H.~W.}:
\newblock A path space extension for robust light transport simulation.
\newblock \emph{ACM Transactions on Graphics (TOG) 31}, 6 (2012), 191.

\bibitem[ICG86]{immel1986radiosity}
\textsc{Immel D.~S., Cohen M.~F., Greenberg D.~P.}:
\newblock A radiosity method for non-diffuse environments.
\newblock \emph{Acm Siggraph Computer Graphics 20}, 4 (1986), 133--142.

\bibitem[JC95]{jc95}
\textsc{Jensen H.~W., Christensen N.}:
\newblock Photon maps in bidirectional monte carlo ray tracing of complex
  objects.
\newblock \emph{Computers and Graphics 19}, 2 (1995), 215--224.

\bibitem[Jen96]{jensen1996global}
\textsc{Jensen H.~W.}:
\newblock Global illumination using photon maps.
\newblock In \emph{Rendering Techniques’ 96}. Springer, 1996, pp.~21--30.

\bibitem[JRJ11]{jakob2011progressive}
\textsc{Jakob W., Regg C., Jarosz W.}:
\newblock Progressive expectation-maximization for hierarchical volumetric
  photon mapping.
\newblock In \emph{Computer Graphics Forum} (2011), vol.~30, Wiley Online
  Library, pp.~1287--1297.

\bibitem[Kaj86]{kajiya1986rendering}
\textsc{Kajiya J.~T.}:
\newblock The rendering equation.
\newblock In \emph{ACM SIGGRAPH computer graphics} (1986), vol.~20, ACM,
  pp.~143--150.

\bibitem[KBS15]{kalantari2015machine}
\textsc{Kalantari N.~K., Bako S., Sen P.}:
\newblock A machine learning approach for filtering monte carlo noise.
\newblock \emph{ACM Trans. Graph. 34}, 4 (2015), 122--1.

\bibitem[KD13]{kaplanyan2013adaptive}
\textsc{Kaplanyan A.~S., Dachsbacher C.}:
\newblock Adaptive progressive photon mapping.
\newblock \emph{ACM Transactions on Graphics (TOG) 32}, 2 (2013), 16.

\bibitem[KR17]{kalantari2017deep}
\textsc{Kalantari N.~K., Ramamoorthi R.}:
\newblock Deep high dynamic range imaging of dynamic scenes.
\newblock \emph{ACM Trans. Graph. 36}, 4 (2017), 144--1.

\bibitem[KWX{\etalchar{*}}16]{kang2016adaptive}
\textsc{Kang C.-M., Wang L., Xu Y.-N., Meng X.-X., Song Y.-J.}:
\newblock Adaptive photon mapping based on gradient.
\newblock \emph{Journal of Computer Science and Technology 31}, 1 (2016),
  217--224.

\bibitem[KZ11]{knaus2011progressive}
\textsc{Knaus C., Zwicker M.}:
\newblock Progressive photon mapping: A probabilistic approach.
\newblock \emph{ACM Transactions on Graphics (TOG) 30}, 3 (2011), 25.

\bibitem[LW93]{lafortune1993bi}
\textsc{Lafortune E.~P., Willems Y.}:
\newblock Bi-directional path tracing.
\newblock In \emph{Compugraphics' 93} (1993), pp.~145--153.

\bibitem[ODR09]{overbeck2009adaptive}
\textsc{Overbeck R.~S., Donner C., Ramamoorthi R.}:
\newblock Adaptive wavelet rendering.
\newblock \emph{ACM Trans. Graph. 28}, 5 (2009), 140.

\bibitem[PKK00]{pauly2000metropolis}
\textsc{Pauly M., Kollig T., Keller A.}:
\newblock Metropolis light transport for participating media.
\newblock In \emph{Rendering Techniques 2000}. Springer, 2000, pp.~11--22.

\bibitem[QSMG17]{qi2017pointnet}
\textsc{Qi C.~R., Su H., Mo K., Guibas L.~J.}:
\newblock Pointnet: Deep learning on point sets for 3d classification and
  segmentation.
\newblock In \emph{Proceedings of the IEEE Conference on Computer Vision and
  Pattern Recognition} (2017), pp.~652--660.

\bibitem[RMZ13]{rousselle2013robust}
\textsc{Rousselle F., Manzi M., Zwicker M.}:
\newblock Robust denoising using feature and color information.
\newblock In \emph{Computer Graphics Forum} (2013), vol.~32, Wiley Online
  Library, pp.~121--130.

\bibitem[SFES07]{schjoth2007photon}
\textsc{Schj{\o}th L., Frisvad J.~R., Erleben K., Sporring J.}:
\newblock Photon differentials.
\newblock In \emph{Proceedings of the 5th international conference on Computer
  graphics and interactive techniques in Australia and Southeast Asia} (2007),
  pp.~179--186.

\bibitem[SJ09]{spencer2009into}
\textsc{Spencer B., Jones M.~W.}:
\newblock Into the blue: Better caustics through photon relaxation.
\newblock In \emph{Computer Graphics Forum} (2009), vol.~28, Wiley Online
  Library, pp.~319--328.

\bibitem[SJ13a]{spencer2013photon}
\textsc{Spencer B., Jones M.~W.}:
\newblock Photon parameterisation for robust relaxation constraints.
\newblock In \emph{Computer Graphics Forum} (2013), vol.~32, Wiley Online
  Library, pp.~83--92.

\bibitem[SJ13b]{spencer2013progressive}
\textsc{Spencer B., Jones M.~W.}:
\newblock Progressive photon relaxation.
\newblock \emph{ACM Transactions on Graphics (TOG) 32}, 1 (2013), 1--11.

\bibitem[SSFO08]{schjoth2008diffusion}
\textsc{Schj{\o}th L., Sporring J., Fogh~Olsen O.}:
\newblock Diffusion based photon mapping.
\newblock In \emph{Computer Graphics Forum} (2008), vol.~27, Wiley Online
  Library, pp.~2114--2127.

\bibitem[SWH{\etalchar{*}}95]{shirley1995global}
\textsc{Shirley P., Wade B., Hubbard P.~M., Zareski D., Walter B., Greenberg
  D.~P.}:
\newblock Global illumination via density-estimation.
\newblock In \emph{Rendering Techniques’ 95}. Springer, 1995, pp.~219--230.

\bibitem[Vea97]{veach1997robust}
\textsc{Veach E.}:
\newblock \emph{Robust Monte Carlo methods for light transport simulation},
  vol.~1610.
\newblock Stanford University PhD thesis, 1997.

\bibitem[VG95]{veach1995optimally}
\textsc{Veach E., Guibas L.~J.}:
\newblock Optimally combining sampling techniques for monte carlo rendering.
\newblock In \emph{Proceedings of the 22nd annual conference on Computer
  graphics and interactive techniques} (1995), ACM, pp.~419--428.

\bibitem[Vor11]{vorba2011bidirectional}
\textsc{Vorba J.}:
\newblock Bidirectional photon mapping.
\newblock In \emph{Proc. of the Central European Seminar on Computer Graphics
  (CESCG'11)} (2011).

\bibitem[VRM{\etalchar{*}}18]{vogels2018denoising}
\textsc{Vogels T., Rousselle F., McWilliams B., R{\"o}thlin G., Harvill A.,
  Adler D., Meyer M., Nov{\'a}k J.}:
\newblock Denoising with kernel prediction and asymmetric loss functions.
\newblock \emph{ACM Transactions on Graphics (TOG) 37}, 4 (2018), 124.

\bibitem[WHSG97]{walter1997global}
\textsc{Walter B., Hubbard P.~M., Shirley P., Greenberg D.~P.}:
\newblock Global illumination using local linear density estimation.
\newblock \emph{ACM Transactions on Graphics (TOG) 16}, 3 (1997), 217--259.

\bibitem[WJ94]{wand1994kernel}
\textsc{Wand M.~P., Jones M.~C.}:
\newblock \emph{Kernel smoothing}.
\newblock Chapman and Hall/CRC, 1994.

\bibitem[WYKR17]{wu2017multiple}
\textsc{Wu L., Yan L.-Q., Kuznetsov A., Ramamoorthi R.}:
\newblock Multiple axis-aligned filters for rendering of combined distribution
  effects.
\newblock In \emph{Computer Graphics Forum} (2017), vol.~36, Wiley Online
  Library, pp.~155--166.

\bibitem[XBS{\etalchar{*}}19]{xu2019deep}
\textsc{Xu Z., Bi S., Sunkavalli K., Hadap S., Su H., Ramamoorthi R.}:
\newblock Deep view synthesis from sparse photometric images.
\newblock \emph{ACM Transactions on Graphics (TOG) 38}, 4 (2019), 1--13.

\bibitem[XSHR18]{xu2018deep}
\textsc{Xu Z., Sunkavalli K., Hadap S., Ramamoorthi R.}:
\newblock Deep image-based relighting from optimal sparse samples.
\newblock \emph{ACM Transactions on Graphics (TOG) 37}, 4 (2018), 126.

\bibitem[XZW{\etalchar{*}}19]{xu2019adversarial}
\textsc{Xu B., Zhang J., Wang R., Xu K., Yang Y.-L., Tang R.}:
\newblock Adversarial {M}onte {C}arlo denoising with conditioned auxiliary
  feature modulation.
\newblock \emph{ACM Transactions on Graphics (TOG) 38}, 6 (2019), 224.

\bibitem[YMRD15]{yan2015fast}
\textsc{Yan L.-Q., Mehta S.~U., Ramamoorthi R., Durand F.}:
\newblock Fast 4d sheared filtering for interactive rendering of distribution
  effects.
\newblock \emph{ACM Transactions on Graphics (TOG) 35}, 1 (2015), 7.

\bibitem[ZJL{\etalchar{*}}15]{zwicker2015recent}
\textsc{Zwicker M., Jarosz W., Lehtinen J., Moon B., Ramamoorthi R., Rousselle
  F., Sen P., Soler C., Yoon S.-E.}:
\newblock Recent advances in adaptive sampling and reconstruction for monte
  carlo rendering.
\newblock In \emph{Computer Graphics Forum} (2015), vol.~34, Wiley Online
  Library, pp.~667--681.

\end{thebibliography}

% biblatex with biber
%\printbibliography                

%-------------------------------------------------------------------------

\end{document}